\documentclass[journal]{IEEEtran}
\IEEEoverridecommandlockouts
\usepackage{cite}
\usepackage{amsmath,amssymb,amsfonts}
\usepackage{graphicx}
\usepackage{textcomp}
\usepackage[dvipsnames]{xcolor}

\usepackage{caption} %
\usepackage{subcaption} %
\usepackage{hyperref}

\usepackage[ruled,vlined]{algorithm2e}
\usepackage{algpseudocode}

\usepackage{bbm}
\usepackage{enumerate}

\usepackage{mathtools}

\usepackage[normalem]{ulem}

\def\BibTeX{{\rm B\kern-.05em{\sc i\kern-.025em b}\kern-.08em
    T\kern-.1667em\lower.7ex\hbox{E}\kern-.125emX}}

\definecolor{lightgreen}{rgb}{0.5,0.7,0.5}

\begin{document}
\bstctlcite{IEEEexample:BSTcontrol}

\title{Balancing Wildfire Risk and Power Outages through Optimized Power Shut-Offs
}
\author{Noah Rhodes, Lewis Ntaimo and Line Roald*
\thanks{
*Corrersponding author. Email: roald@wisc.edu

Noah Rhodes and L. Roald are with the University of Wisconsin, Madison. 

Lewis Ntaimo is with Texas A\&M University.
}
}

\maketitle

\begin{abstract}
 Electric grid faults can cause catastrophic wildfires, particularly in regions with high winds and low humidity. In real-time operations, electric utilities are often left with few options for wildfire risk mitigation, leading to use of disruptive measures such as proactive de-energization of equipment, frequently referred to as public safety power shut-offs. Such power shut-offs have significant impacts on customers, who experience power cuts in an attempt to protect them from fires. This work proposes the optimal power shut-off problem, an optimization model to support short-term operational decision making in the context of extreme wildfire risk. Specifically, the model optimizes grid operations to maximize the amount of power that can be delivered, while proactively minimizing the risk of wildfire ignitions by selectively de-energizing components in the grid. This is the first optimization model to consider how preventive wildfire risk measures impact both wildfire risk \emph{and} power systems reliability at a short-term, operational time-frame. The  effectiveness of the method is demonstrated on an augmented version of the RTS-GMLC test case, located in Southern California, and compared against two approaches based on simple risk thresholds. The proposed optimization-based model reduces both wildfire risk and lost load shed relative to the benchmarks. 
\end{abstract}

\section{Introduction}
A number of tragic wildfires in recent years have highlighted the loss of life and property that may originate from electric faults. 
In Victoria, the 2009 Black Saturday wildfires killed 179 people. Several of those fires, including the most deadly, were sparked by electric power infrastructures \cite{victoria2009final}.
In Texas, two 2011 wildfires in Bastrop county started by trees coming in contact with nearby power lines \cite{Texas-Bastrop}, and became Texas's most destructive wildfires in history, killing 4 and causing more than \$300 million in damage.
In California, the 2018 Camp Fire, which was sparked by a power line, killed 84 people and caused an estimated \$9.3 billion in residential property damage alone \cite{CoreLogic-Wildfire}. This and other fires ignited during the 2017 and 2018 California fire seasons lead the responsible utility Pacific Gas \& Electricity (PG\&E) to file for bankruptcy \cite{PGE-bankruptcypressrelease} and accept charges for involuntary manslaughter \cite{PGE-manslaughterpressrelease}. %

The risk of wildfire ignitions by power system infrastructure is exacerbated by the fact that power failures are more likely to occur during windy conditions, when wildfires spread faster and are harder to contain.
As a result, research on Australian bushfires and ignition sources in California has found that fires ignited by power lines tend to be larger and more damaging than others \cite{victoria2009final, miller2017electrically, keeley2018historical}. Ignitions caused by power lines are not uncommon -- in Texas, it is estimated that electric equipment caused more than 4,000 fires in less than 4 years \cite{TAMU-wildfiremitigation}, while  PG\&E reported 414 ignition events from 2015-17 \cite{PGE-amendedsafetyplan}. 

Power infrastructure cause ignitions in a number of ways \cite{russell2012distribution, benner2019dfa,jazebi2020}, with the most common cause being contact between vegetation and conductors. Efforts to reduce probability of ignitions include increased frequency of inspections, more aggressive vegetation management, and changes to the protection systems to reduce the number of reclosing attempts or limit the fault current \cite{PGE-amendedsafetyplan, SCE-safetyplan, PGE-2020safetyplan, jazebi2020} .
However, inspections, vegetation management, and equipment upgrades must be planned over a seasonal or yearly time-scale. In day-to-day operations, utilities are left with fewer and more disruptive actions to reduce wildfire risk. 
One measure available to utilities in this shorter time-frame is to limit or turn off automatic reclosing to ensure that a fault location does not experience multiple arcing events \cite{SCE-safetyplan, PGE-amendedsafetyplan}. However, this does not mitigate ignitions caused by the initial fault, nor does it avoid ignitions from high-impedance distribution line faults which may go undetected \cite{russell2012distribution}. 
For safety during extreme conditions, the only measure that completely removes risk of fire ignitions is therefore to completely turn off a power line, as a de-energized line will not cause any sparks. %
After the deadly and devastating fires caused by electric power lines in 2017 and 2018, %
California utilities expanded the use of intentional de-energization of power lines during high-risk conditions to avoid faults that may spark new fires.

While de-energization is arguably the only way to completely avoid the risk of ignitions, it also impacts the ability of the power system to provide reliable electricity. At their peak in October 2019, the intentional blackouts that result from de-energization, typically referred to as public power safety shut-offs (PSPS) \cite{PGE-amendedsafetyplan, SCE-safetyplan}, turned off power to almost a million customer accounts \cite{CPUC-Deenergization}. The economic and societal impacts of a blackout on this scale, lasting for several days, is enormous, and also has important impacts on health such as increased mortality \cite{anderson2020}. %
After the first major power shut-off in northern California, PG\&E confirmed no less than 100 incidents of wind damage (including downed lines and contact with trees) along the 25,000 miles of power lines that were taken out of service \cite{PGE-Opinion}. Any of those incidents could have sparked a fire.

\emph{Given the stakes, the risk of wildfires versus the disruption of power outages, it is a challenging, but critically important task to decide which power equipment to turn off during high-risk conditions}. Statements by PG\&E regarding the 2019 Kincade fire, which was likely ignited by a high-voltage transmission in an area where all distribution lines were de-energized, indicate that current operational procedures mostly consider wind speed, wildfire risk, and voltage levels when deciding which lines to turn off \cite{Kincade}. This raises the question of whether a more targeted approach could achieve better results by considering the risk of individual components, while accounting for both the wildfire risk reduction and the reduction in electric load delivery associated with turning a specific component off. The need for less indiscriminate use of public safety power shut-offs has also been highlighted by the California government \cite{newsom}. This paper presents a first step towards developing a more targeted model, and demonstrates that \emph{proactively considering both the risk of wildfires and the impact of a power outages when optimizing power system operations allows for a better balance between the two.}

Previous work on system operations under wildfire risk is surprisingly limited. %
Existing work has focused primarily on the impact of fires on the operation of power system equipment, such as the risk of wildfire exposure for transmission lines  \cite{sathaye2013rising, dian2019integrating}, or operational impacts due to reductions in the transmission line thermal capacity as a result of heat from a nearby fire \cite{choobineh2016power, trakas2017optimal}. While the existing work considers the challenge of operating the grid in presence of fires, this paper focuses on how power system operators can take an active role in \emph{preventing} fires from occurring. Specifically, we provide a modeling and optimization framework to support the decision process surrounding a public safety power shut-off, with the aim of maintaining as much load delivery as possible while minimizing wildfire risk. To the best of our knowledge, this is the first paper to describe such a model.  

The contributions of our paper are threefold.\\
(1) We review existing methods for wildfire risk assessment, and propose a method to  estimate the relative risk associated with specific electric components. This allows us to assess the wildfire risk reduction associated with de-energizing  this component.\\
(2) We propose an optimization framework which minimizes wildfire risk, while maximizing the amount of load that can be delivered. This optimization is based on a DC power flow representation, and uses a risk parameter $\alpha$ to weight the risk of wildfire vs the willingness to shed load.\\
(3) We compare our method against two benchmarks on the RTS-GMLC test case, which was augmented to include data on wildfire risk. The case study demonstrates the value of accounting for both wildfire risk and impact on load shedding when determining which components to de-energize. 

The remainder of the paper is organized as follows. Section \ref{sec:wildfire} describes interactions between wildfire risk and electric grid operation, while Section \ref{sec:optimization} provides details on the optimization problem formulation. Section \ref{sec:benchmarks} describes the benchmark approaches, while Sections \ref{sec:case1} and \ref{sec:case2} describes the case study set-up and numerical results. Section \ref{sec:conclusions} summarizes and concludes, and gives an outline of future work.

\section{Wildfire Risk and Electric Grids}
\label{sec:wildfire}
The main mechanism for fire ignitions by electric power lines are electric faults, which cause sparks or arcing in a high-heat release of current \cite{benner2019dfa}. 
The majority of such ignitions are caused by vegetation coming into contact with power lines \cite{PGE-amendedsafetyplan}.
This leads to an unfortunate correlation between wildfire risk and electric power failures. Extreme fire conditions are characterized by access to fuel, low humidity and high wind conditions. Examples of highly flammable fuels include grasslands, bushes or dead trees \cite{calfire2020}. Low humidity leads to a high probability that an ignition will lead to a fire, particularly in regions with fire-prone vegetation. High wind conditions imply that fires spread faster and are harder to contain. 

At the same time, high wind conditions cause increased movements of power lines and nearby trees, and hence increase the probability of power line faults and subsequent ignitions.
This correlation between decreased ability to contain a fire and increased probability of ignitions from power grid infrastructure is what motivates the use of proactive de-energization of equipment as a preventive, though disruptive measure. 
In the following, we first present our model for the wildfire risk associated with electric components, which allows us to quantify the risk reduction achieved through equipment de-energization. We then discuss the trade-off between reducing the risk and maintaining electric load delivery.

\subsection{Modelling Risk of Wildfire Ignitions} 

Modelling wildfire risk and its relationship with the operation of electric power system is highly challenging. 
Generally speaking, wildfire risk can be defined as the product of the probability or likelihood of a fire occurring and the impact of the fire when it occurs. To make this definition and the relationship to electric power systems more concrete, we start with a review of the wildfire risk literature, where we have identified four main components to wildfire risk.
\begin{enumerate}[(i)]
    \item \textbf{Probability of an ignition.} Ignitions are the initiating event for a fire and are a localized effect that happen within a given area. There may be ignitions from several sources, with ignition probabilities from different sources varying over the course of the year \cite{syphard2015location,keeley2018historical}. 
    \item \textbf{Probability that an ignition leads to a fire.} The probability that an ignition grows to become a (large) fire is dependent on the immediate local vegetation, humidity and topology. Measures such as the fire potential index captures the susceptibility of an area to an ignition \cite{USGS-Fire}, while the large fire probability  captures how likely it is that a small fire of one acre would grow beyond 100 acres in size \cite{USGS-EROS}.
    \item \textbf{The size and intensity of the resulting fire.} The size and intensity of the fire is also determined by local vegetation, humidity, topology and weather conditions. This is referred to as fire behavior, and can be modelled using simulation tools such as FlamMap \cite{flammap}.%
    \item \textbf{Damage caused by the fire.} The damage caused by a fire is related to the value of resources inside the area, as well as the fire intensity. %
    Large-scale, high-intensity fires are typically associated with substantial damage to property, resources and loss of life, and therefore have a high negative impact. On the other hand, (low-intensity) fires are a natural and necessary component in many ecosystems and may also have a positive impact \cite{scott2013wildfire}. 
\end{enumerate}

\subsubsection{Traditional Wildfire Risk Assessment}
Traditional assessments of wildfire risk capture these four aspects to a varying degree. 
For example, the \emph{likelihood} that a given area experiences a fire incorporates both (i) and (ii), and can be combined with studies of the \emph{fire intensity} (iii) to determine the so-called \emph{wildfire threat} \cite{finney2010continental}. Wildfire threat data is often provided at a pre-defined granularity, with areas ranging in size from 30 by 30 meters \cite{Texas-WRAP} to one square kilometer \cite{USGS-Fire}. 
Examples of wildfire risk maps that consider wildfire threat include California Fire and Resource Assessment Program's Fire Hazard Severity Zone (FHSZ) \cite{CPUC-FRAP}, where the threat level is reported on a scale from 1 to 5 to show the relative risk in various regions. The United States Forest Service releases the Wildfire Hazard Potential \cite{dillon2015wildland} with relative values ranging from low to high. %
The Texas Wildfire Risk Assessment System (TWRAS) \cite{arrubla2014wildfire, Texas-WRAP} combines assessments of wildfire threat with information about the damage incurred by a fire through the \textit{Wildland-Urban Interface Response Index}, which reports the net impact on housing resources and human life based on wildfire intensity in a region. The response index is a net value change on the resources, ranging from -1 (least negative impact) to -9 (most negative impact).  
 
It important to note that all these sources present wildfire risk at an approximate, \textit{relative} scale, showing regions that have a greatly elevated fire potential, wildfire threat or where significant resources are at risk of fire damage.
The scale is used to aid forest managers, fire departments and local governments in how to effectively respond to a wildfire threat through measures such as prescribed burns or strategic positioning of fire fighting resources \cite{arrubla2014wildfire}. 
Thus, existing wildfire risk assessments are comprehensive and well-designed for long term fire risk planning which reduces the probability that a fire develops and limit the size and intensity of a fire. However, because these existing assessment tools frequently only implicitly account for ignition sources, they must be adapted to assess the wildfire risk mitigation achieved by reducing the probability of ignitions from electric grid infrastructure. 

\subsubsection{Wildfire Risk of Electric Components}
Considering the four components of wildfire risk (i) - (iv), we observe that the only component that can be controlled by the electric utility (at least at a shorter time-frame) is the probability of ignition events (i). We therefore discuss our approach to model the impact of a reduction in the ignition probability. 

Assume that we are provided a wildfire map with pre-defined areas $j\in\mathcal{A}$, where each area is associated with a wildfire risk value $R_j$. This value may represent for example the wildfire threat \cite{finney2010continental, dillon2015wildland, CPUC-FRAP} or include information about values at risk of fire damage \cite{Texas-WRAP}.

Next, we make an assumption about the ignition sources. While there are many different sources of ignitions, it is known that electric power infrastructure are more likely to fail during certain weather conditions and are the most likely cause of fires during certain parts of the year \cite{syphard2015location}. For the purposes of our analysis, we therefore assume that high-voltage electric power equipment represent the dominant source of ignitions in their immediate vicinity (i.e., in the area where they are located) and are responsible for the full wildfire risk $R_j$. 

However, the probability of an ignition from electric power equipment in a given area is directly related to operational decisions such as de-energization (which brings this probability to zero), and is also impacted by equipment characteristics such as the voltage level of power lines, maintenance status or recent vegetation management. For example, if a line has a higher fault rate than average or maintenance is overdue, the ignition rate can be increased. Similarly, lower voltage transmission lines have a narrower right-of-way, and are more likely to have vegetation blown into them by strong winds. Thus, the relative ignition rate of lower voltage lines may be set higher, particularly if vegetation management is lagging behind.
This leads to relative variations in the ignition probabilities for different parts of the power systems, which are not accounted for in the standard wildfire risk assessments provided by government agencies as they are only known to the grid operator. We assume that the grid operator utilizes this information to enhance the wildfire risk model by incorporating a relative risk factor $\kappa_{e,j}$, which can be used to represent an elevated or reduced risk of ignitions from the electric component $e$ in area $j$. This gives rise to the following risk model for the risk $R_{e,j}$,
\begin{equation}
    R_{e,j} = \kappa_{e,j} R_j.
\end{equation}
Here, $\kappa=1.0$ is used as a baseline value for the probability of ignitions, while $\kappa > 1.0$  and $\kappa < 1.0$ indicate elevated or reduced risk of an ignition from an electric component. {\color{black}{The value of $\kappa$ can be determined by a utility  based on historical data on past fault locations, fault conditions and ignitions, current system data including time since the last tree trimming, equipment age and maintenance, and weather forecasts including wind speeds. For example, a transmission line which has experienced multiple faults in the past, is going through a wooded, windy area and has not been subject to recent tree trimming presents an elevated risk with $\kappa>1.0$. On the other hand, an underground cable (which essentially has zero risk of igniting nearby vegetation) would have zero risk $\kappa=0$. %
}}
For components that stretch across multiple areas $j\in\mathcal{A}_e$, where $\mathcal{A}_e\subset\mathcal{A}$ is the set of areas that component $j$ passes through, the total risk is the cumulative risk across all areas and is computed by taking the sum,
\begin{equation}
    \boldsymbol{R}_{e} =  \sum_{j\in \mathcal{A}_e} \kappa_{e,j} R_{j} .
    \label{eq:multiarea}
\end{equation}
We note that the risk model (2) does not capture the impact of a fire that spreads from area $j$ to adjacent areas and thus is only a local measure of wildfire risk. However, we believe this is a reasonable model as the electric utility bears responsibility for avoiding wildfires in the vicinity of their equipment, but does not have control over mitigation measures to limit the spread once a fire is ignited such as large scale fuel management or dispatch of fire fighting services.

\subsubsection{Short-Term Mitigation of Wildfire Risk}
While long-term planning to reduce wildfire risk include measures such as increased inspections or more aggressive vegetation management around power lines, these mitigation measures take time to plan for and implement.
In short-term operations, operators therefore have fewer options, with the most effective (and most disruptive) being de-energization of equipment. If a component is de-energized, risk is reduced to zero, as it will not cause any faults or ignitions.
The decision on whether to keep a component energized can hence be understood as a  decision related to whether or not we are willing to accept the risk $\boldsymbol{R}_e$. With this consideration, we express the wildfire risk as
\begin{equation}
    \boldsymbol{R}_e = \left\{\begin{matrix*}[l] 
     ~\sum_{j\in \mathcal{A}_e} \kappa_{e,j} R_j & \text{ if $e$ is energized} \\ 
     ~0 & \text{ if $e$ is de-energized}
    \end{matrix*}   \right. .
\end{equation}

For generators, lines, and buses, we represent the decision of whether or not to de-energize the equipment through binary decision variables $z_*\in\{0,1\}$ that indicate whether a certain component is energized or not, with $z_g, z_l$, and $z_i$ denoting the variables for generators, lines, and buses, respectively. In the case of loads, we assume that the load seen by the operator is representative of a large number of individual loads. We therefore model load shedding as a continuous variable $x_d\in [0,1]$  that represent the fraction of the load (and corresponding distribution infrastructure) that is de-energized.

This allows us to model $R_{Fire}$, the total wildfire risk arising from electric components, as
\begin{equation}
    {R_{Fire}} = \sum_{d\in\mathcal{D}} x_d \boldsymbol{R}_d +  \sum_{g\in\mathcal{G}}z_g \boldsymbol{R}_g + \sum_{l\in\mathcal{L}}z_l \boldsymbol{R}_l + \sum_{i\in\mathcal{B}}z_i \boldsymbol{R}_i,  \label{eq:risktot}
\end{equation}
where $\mathcal{D}, \mathcal{G}, \mathcal{L}$, and $\mathcal{B}$ represent the sets of load demand, generators, transmission lines, and buses, respectively. 
We note that the units of wildfire risk will be determined by the units of wildfire risk map we are using. Since these maps are often represented in an approximate, relative scale (such as \emph{low} or \emph{high}), the risk of the electric components would be ranked on a similarly coarse scale. We note that this course assessment is also a reflection of the challenges associated with assigning detailed numerical values to the economic and societal impacts of fire, including loss of life. 

\subsection{Modelling Risk of Power Shut-Offs}
De-energizing loads and electric equipment not only reduces wildfire risk, but also limits the ability of the system to provide electricity to customers, leading to power shut-offs.
Similar to the wildfire risk modelling, it is very hard to obtain detailed and accurate estimates of the economic and societal impact of such power shut-offs. The cost depends on a number of factors, including the frequency and duration of such shut-offs, as well as the value the electricity provides to individual customers. %
We thus refrain from a detailed monetary assessment of the economic impact of lost load, and aim to deliver as much load as possible. Our variable $D_{Tot}$ represents the total amount of load delivered, and is expressed by
\begin{equation}
    {D_{Tot}} = \sum_{d\in\mathcal{D}} x_d w_d \boldsymbol{D}_d, \label{eq:dtot}
\end{equation}
where $\boldsymbol{D}_d$ is the amount of load served under normal operating conditions. 
We include the weight $w_d$ to express that certain loads, such as hospitals or other essential services, may be prioritized over others. 

\subsection{Trading off Risk of Wildfires and Power Shut-Offs}
With the above modelling, we express the trade-off between maximizing load delivery and minimizing wildfire risk as 
\begin{equation}
\label{eq:obj}
    \max \quad (1-\boldsymbol{\alpha}){D_{Tot}} - \boldsymbol{\alpha} {R_{Fire}}
\end{equation}
Here, the parameter $\alpha \in [0,1]$ expresses the trade-off between serving more load (low $\alpha$) and avoiding wildfire risk (high $\alpha$). 

We note that although the approximations of the wildfire risk associated with each component may seem crude and potentially very conservative, the absolute risk values in the above model are less important than the relative differences in risk among different electric grid components. Furthermore, by solving the model for different values of $\alpha$, we obtain a Pareto front of solutions that represent optimal trade-offs between continued load delivery and reductions in wildfire risk. Because we solve the Pareto front for all possible values of $\alpha$, the units of lost load and wildfire risk do not need to be the same.  We do not prescribe a single optimal solution to the tradeoff, but rather provide a range of possible solutions for the operator to choose from based on their preference.

\section{Optimal Power Shut-Offs}
\label{sec:optimization}

Based on the above modelling considerations, we now present the optimal power shut-off problem, an optimization framework to aid decisions related to public safety power shut-offs. 
We approach the optimization problem from the perspective of a power system operator, whose main objective is to maximize load delivery while minimizing wildfire risk as expressed by \eqref{eq:obj}. The time frame of this decision is making short-term operations, ranging from a few days ahead until real-time operation. We assume that the operator would use this method as part of day-to-day operations to make decisions on where and when to shut off power throughout periods with elevated wildfire risk. The decision variables are $z_g, z_l, z_i$ and $x_d$, representing whether or not individual components will remain energized or not, as well as the power generation of the generators $P_g^G,~g\in \mathcal{G}$, the power flows on transmission lines $l\in\mathcal{L}$ from bus $j$ to bus $i$ given by $P_{l,i,j}^L$ and the voltage angles $\theta_i,~i\in\mathcal{B}$. 
In the following, we present the constraints of our model, and summarize the full optimization problem.

\subsubsection{Component interactions}
Generators, loads and lines can only be energized if the buses they are connected to are energized. We enforce this through the following constraints,
\begin{subequations}
\label{eq:relationships}
\begin{align}
& \quad z_i \ge x_d && \forall d \in \mathcal{B}^\mathcal{D}_i~, \quad \forall i \in \mathcal{B}~,  \label{eq:gen_active} \\
& \quad z_i \ge z_g && \forall g \in \mathcal{B}^\mathcal{G}_i~~, \quad \forall i \in \mathcal{B}~,   \label{eq:line_active}\\
& \quad z_i \ge z_l &&\forall l \in \mathcal{B}_i^{\mathcal{L}}~, \quad \forall i \in \mathcal{B}~.   \label{eq:demand_active}
\end{align}
\end{subequations}
The sets $\mathcal{B}^{\mathcal{D}}_i$, $\mathcal{B}^{\mathcal{G}}_i$, and $\mathcal{B}^{\mathcal{L}}_i$ represent the set of loads, generators, and lines connected to bus $i$, respectively. %

\subsubsection{Generators}
The generator limits are given by
\begin{equation}
    \quad z_g\underline{\boldsymbol{P}_g} \le P _g^G \le z_g \overline{\boldsymbol{P}_g} \quad \forall g \in \mathcal{G}~, \label{eq:gen_limits}
\end{equation}
which enforces that $P_g=0$ if the generator is turned off, and otherwise enforces the maximum and minimum generation limits denoted by $\overline{\boldsymbol{P}_g}, \underline{\boldsymbol{P}_g}$.

\subsubsection{Power flow representation and  transmission limits}
To represent the power flow in the system, we utilize a DC power flow representation, extended to model the on/off status of loads, generators, lines or buses. This gives rise to the following set of equations:
\begin{subequations}
\label{eq:powerflow}
\begin{align}
& P_{l,i,j}^L \le -\boldsymbol{b}_l (\theta_i - \theta_j + \boldsymbol{\theta}^{max}(1-z_{l})) && \forall l \in \mathcal{B}_{i,j}^\mathcal{L} \label{eq:flow_limit1} \\
& P_{l,i,j}^L \ge -\boldsymbol{b}_l (\theta_i - \theta_j + \boldsymbol{\theta}^{min} (1-z_{l})) && \forall l \in \mathcal{B}_{i,j}^\mathcal{L} \label{eq:flow_limit2}\\
&-\boldsymbol{T_l}z_{l} \le  P_{l,i,j}^L \le \boldsymbol{T_l}z_{l} && \forall l \in \mathcal{L} \label{eq:thermal_limit}\\
& \sum_{g\in\mathcal{B}_i^\mathcal{G}}P_{g}^G + \sum_{l\in\mathcal{B}_i^\mathcal{L}}P_{l,i,j}^L - \sum_{d\in\mathcal{B}_i^\mathcal{D}}x_d \boldsymbol{D}_d = 0 && \forall i \in \mathcal{B}  \label{eq:power_balance} 
\end{align}
\end{subequations}
Eqs. \eqref{eq:flow_limit1}-\eqref{eq:thermal_limit} represent the power flow $P_{l,i,j}$ on line $l$ from bus $i$ to bus $j$, with $\boldsymbol{b}_l$ representing the susceptance of the line. %
We note that if line $l$ is energized, i.e. $z_{l}=1$, \eqref{eq:flow_limit1} and  \eqref{eq:flow_limit2} correspond to the standard DC power flow constraint, and the power flow is limited to be within the thermal power flow limit $\boldsymbol{T_l}$ by \eqref{eq:thermal_limit}. If the line is de-energized, $z_l=0$, we relax \eqref{eq:flow_limit1}, \eqref{eq:flow_limit2} by adding constants $\boldsymbol{\theta^{max}}, \boldsymbol{\theta^{Min}}$ to the angle differences, such that these constraints can never be binding. 
Further, for $z_l=0$, \eqref{eq:thermal_limit} the power flow $P_{l,i,j}$ is set to zero. 
The nodal power balance is represented in equation \eqref{eq:power_balance}, where the variable $x_d$ represents the fraction of the load $\boldsymbol{D}_d$ that is shed.

\subsubsection{Optimal Power Shut-Off Problem}
With the above modeling considerations, the full optimization problem is given by
\begin{subequations} \label{eq:ops}
\begin{align}
    &\max\limits_{\substack{x, z, P^G, P^L, \theta\\ D_{Tot}, R_{Fire}}} &&  \mbox{$(1-\boldsymbol{\alpha}){D_{Tot}} - \boldsymbol{\alpha} {R_{Fire}}$ \eqref{eq:obj}} \nonumber\\
&\mbox{s.t.: \,\,\,}
&& 
\mbox{Component relationship:}~\eqref{eq:relationships} \tag{OPS}\\
&&& \mbox{Generator constraints:}~\eqref{eq:gen_limits} \nonumber\\
&&& \mbox{Power flow constraints:}~\eqref{eq:powerflow} \nonumber
\end{align}
\end{subequations}
We note that this is a mixed-integer linear program (MILP), with integer variables for all components in the system.

\section{Benchmark Methods}
\label{sec:benchmarks} 
 
We compare the optimal power shut-off \eqref{eq:ops} problem against two heuristics which decide which components to de-energize based solely on the associated wildfire risk $R_e$. This is inspired by accounts of the public safety power shut-offs in the area around the Kincade fire \cite{Kincade}. 
These heuristics are implemented as a two-step procedure, where the first step decides which components to turn off by checking the wildfire risk level associated with the components against a pre-defined threshold and the second step maximizes the amount of load that can be served given that the high-risk components have been turned off. 
\subsection{Step 1: Component shut-off}
We consider two heuristics which differ in the way the wildfire risk threshold is implemented.\\
\noindent\emph{1) Area Heuristic (AH)}
The first benchmark computes the overall wildfire risk $R_\mathcal{A}$ within a predefined area of the electric grid. If the total risk is above a given threshold $R_\mathcal{A}^{max}$, all components within this area are shut off.\\[+4pt]
\textbf{Area criterion:} $R_\mathcal{A} = \sum_{e\in\mathcal{A}} R_e \geq R_\mathcal{A}^{max}$\\[+6pt]
\noindent\emph{2) Transmission Heuristic (TH)} 
The second benchmark considers the risk $R_e$ of individual transmission lines.  If a line exceeds the pre-fixed risk threshold $R_e^{max}$, it is deactivated. \\[+4pt]
\textbf{Transmission criterion: } $R_l \geq R_l^{max}$ \\[+4pt]
In this model, we do not directly shut down loads, generators or buses, except when they have to be shut down due to lack of interconnection. \\[+2pt]
For both of these heuristics, we refer to the set of components $c$ that should be de-energized as $c \in \mathcal{C}$. %
\subsection{Step 2: Maximize Load Delivery}
After we have identified which loads that need to be switched off, we solve a version of the maximum load delivery (MLD) problem. This problem formulation was developed to identify the largest load delivery in a situation with large-scale N-k outages \cite{MLD}. When a significant number of components are out of service, the network may contain several islands, some of which may not be able to serve the full load or may have load that is below the minimum generation limit of the generators. As a result, a normal optimal power flow problem may be infeasible, and it is necessary to include variables to allow further components to be disabled, such as isolated generators or additional load shed. This gives rise to the following optimization problem, 
\begin{subequations} \label{eq:mld}
\begin{align}
    &\max\limits_{\substack{x, z, P^G, P^L,\\ \theta, D_{Tot}}} &&  \!\!\!\!\!\!\!\!\!\!\!\!\!\!\!\!\mbox{${D_{Tot}}$ \eqref{eq:dtot}}  \nonumber\\
&\mbox{s.t.: \,\,\,} && \!\!\!\!\!\!\!\!\!\!\!\!\!\!\!\!\mbox{$z_c = 0 \quad \forall c \in  \mathcal{C}$} \tag{MLD} \\
&&& \!\!\!\!\!\!\!\!\!\!\!\!\!\!\!\!\mbox{Component relationship:}~\eqref{eq:relationships} \nonumber\\
&&& \!\!\!\!\!\!\!\!\!\!\!\!\!\!\!\!\mbox{Generator constraints:}~\eqref{eq:gen_limits} \nonumber\\
&&& \!\!\!\!\!\!\!\!\!\!\!\!\!\!\!\!\mbox{Power flow constraints:}~\eqref{eq:powerflow} \nonumber
\end{align}
\end{subequations}
This problem is identical to the OPS problem \eqref{eq:ops}, except for a different objective function and constraints requiring some components to be disabled. The objective function does not consider wildfire risk, and only maximizes the load served (corresponding to $\alpha = 0$ in the OPS problem). 
Each component $c \in \mathcal{C}$ that is disabled by the heuristic has the corresponding energized binary variable $z_c$ constrained to 0 (de-energized). The remaining constraints are the same as in the OPS problem \eqref{eq:ops}.

After we solve the MLD problem, a final post-processing step is used to identify and disable network islands that have no attached load, typically buses that are not connected to either loads, generators or transmission lines. %

We note a few important differences between these heuristics and the OPS model. First, neither of these heuristics considers how turning off the components impacts the ability of the system to serve the load. The decisions on which components are turned on or off are taken solely based on wildfire risk, without consideration of how much load will be shed as a result. 
Second, the optimization problem \eqref{eq:mld} does not consider wildfire risk at all. As a result, some components might be left energized, even if they are not needed to serve the system load. This unnecessarily increases wildfire risk. 
In contrast, the OPS model integrates the two steps into one optimization problem. It is therefore possible to consider the mutual relationship between wildfire risk reduction and load shed achieved by shutting off certain lines.

\section{Case Study: Implementation and Setup}
\label{sec:case1}

\subsection{Implementation}
The Optimal Power-Shutoff Problem and the accompanying heuristics are implemented in the Julia programming language \cite{julia} using JuMP \cite{jump}, and is part of a new package PowerModelsWildfire.jl {
which we have made publicly available in conjunction with this paper\footnote{%
The package PowerModelsWildfire.jl can be found on \url{https://github.com/WISPO-POP/PowerModelsWildfire.jl}}.} This package also relies heavily on the PowerModels.jl packages \cite{coffrin2018powermodels}.
The network plots are generated using a modified version of PowerModelsAnalytics.jl. The maximum load delivery problem \cite{MLD} is solved using PowerModelsRestoration.jl.

Both the optimal power shut-off \eqref{eq:ops} problem and the maximum load delivery (MLD) problem used for the heuristics are mixed-integer linear programs. These problems were all solved in Julia v1.4 using the Gurobi v8.1 optimizer \cite{gurobi} on a machine with a 12-thread Intel CPU @3.2 GHz and 16 GB of memory. With this setup, the optimal power shut-off problem solves in $<0.5$s, while the maximum load delivery problem solves in $<0.05$s.

\subsection{System model with wildfire risk}
{%
To test our models, we create a test case for a power system in an area with high wildfire risk. The test case described in the following is available as part of the PowerModelsWildfire.jl package. }
We start with the RTS-GMLC 73-bus case \cite{8753693} and utilize the standard system configuration as given in the MatPower case file, except that we omit the HVDC line. %
The values for $\boldsymbol{\theta^{max}}$ and $\boldsymbol{\theta^{min}}$ are the maximum possible angle differences that could appear between any two nodes in the system, calculated as described in \cite{hijazi2017convex}. %

\begin{figure}%
    \centering
    \includegraphics[width=0.38\textwidth]{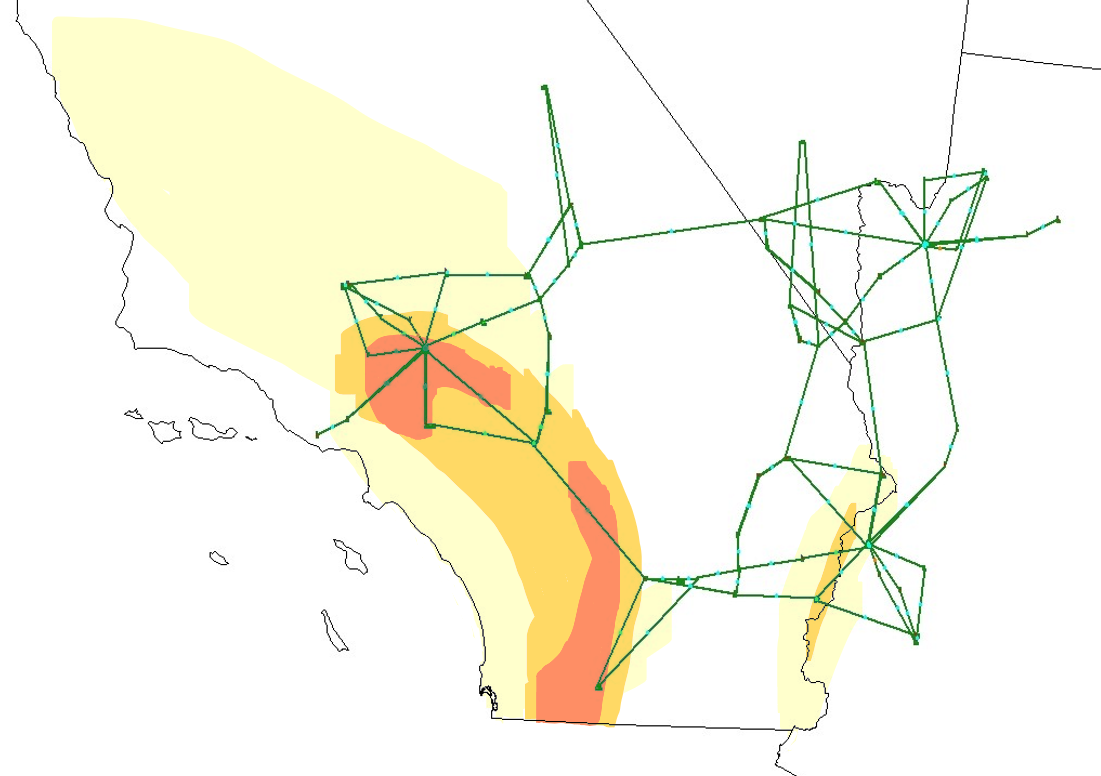}
    \caption{\small \textbf{Wildfire Risk Map} Power grid test case plotted over a representative wildfire risk map for Southern California, based loosely on data in \cite{duginski_2019}. 
    }
    \label{fig:risk_map}
\end{figure}

\begin{figure*}%
    \centering
     \begin{subfigure}[t]{0.3\textwidth}
         \centering
         \includegraphics[width=1.0\textwidth]{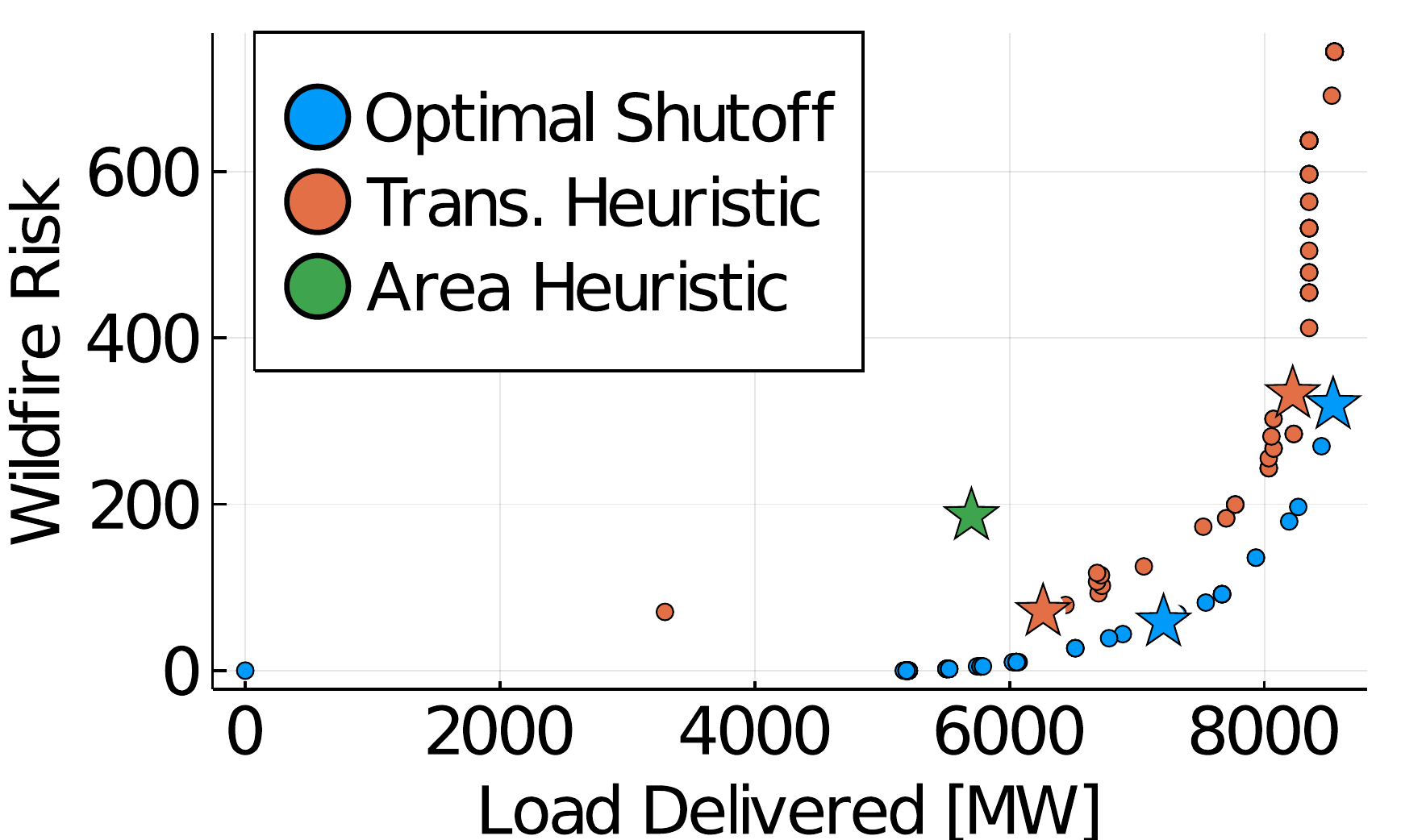}
         \caption{Wildfire risk vs load delivery trade-off curve for all methods.}
         \label{fig:pareto}
     \end{subfigure}
     \hspace{0.08in}
    \begin{subfigure}[t]{0.30\textwidth}
            \centering
            \includegraphics[width=1.0\textwidth]{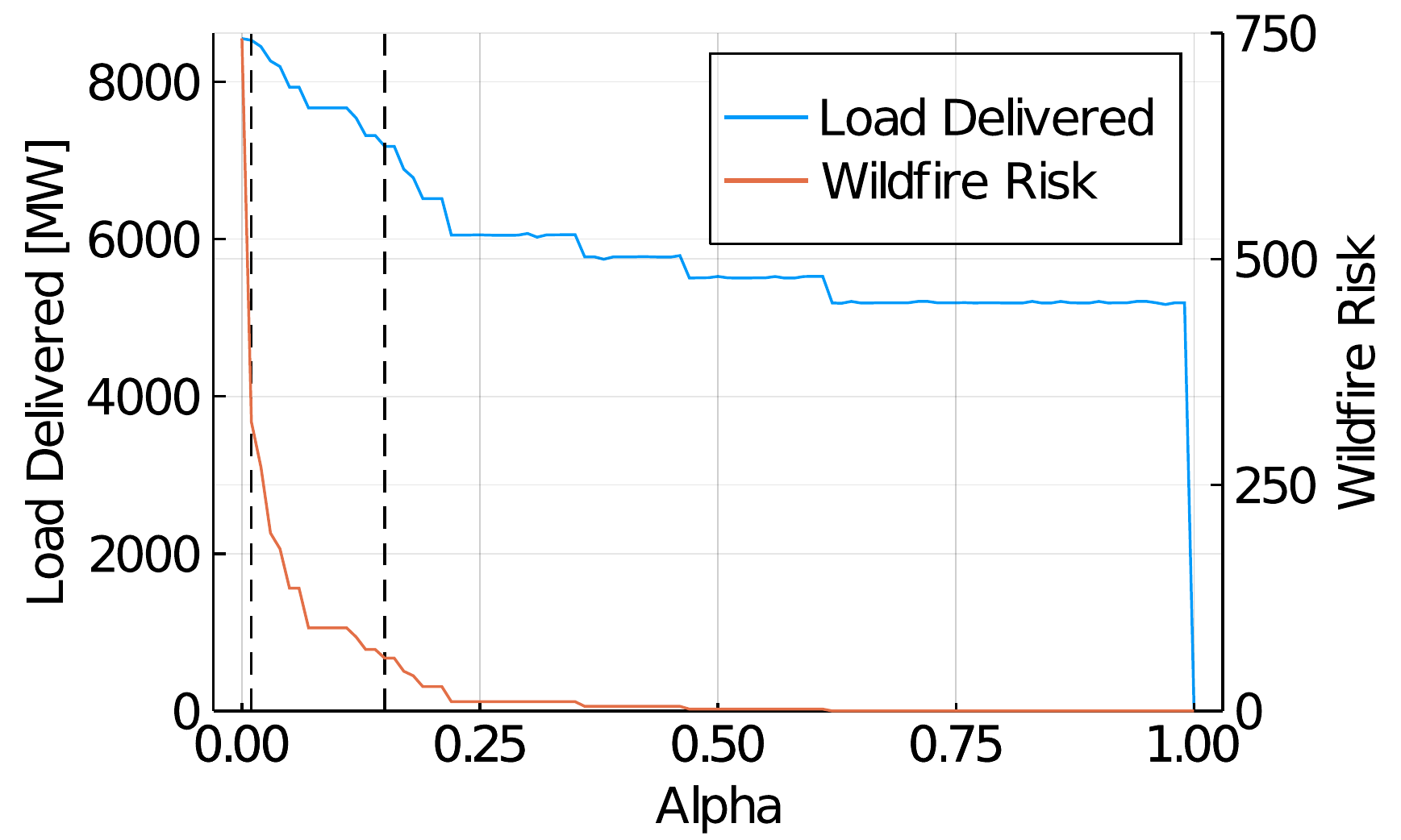}
            \caption{\small Wildfire risk and load delivery as a function of $\alpha$ for the optimal power shut-off. }
            \label{fig:sensitivity_alpha}
     \end{subfigure}
     \hspace{0.08in}
     \begin{subfigure}[t]{0.30\textwidth}
        \centering
        \includegraphics[width=1.0\textwidth]{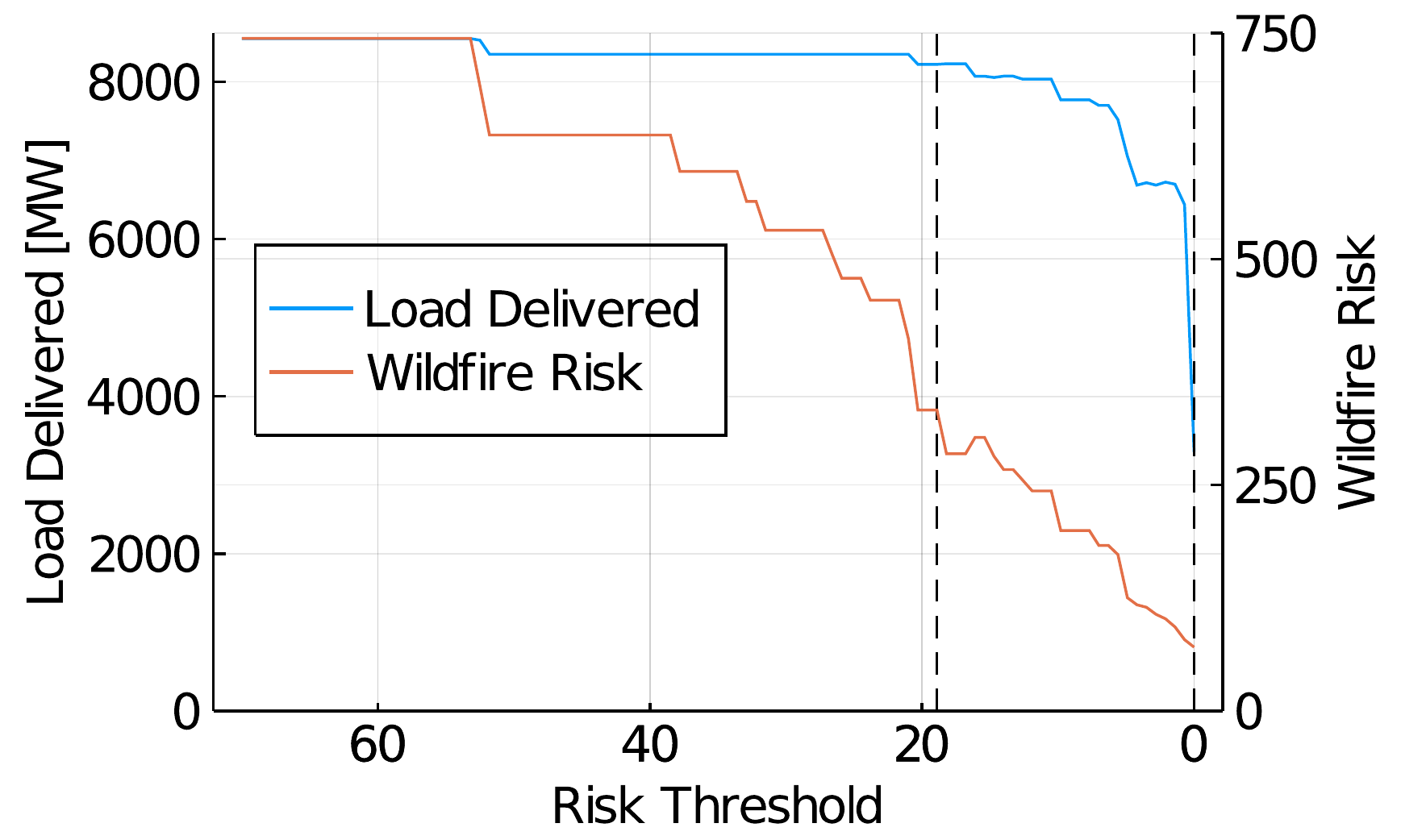}
        \caption{\small Wildfire risk and load delivery as a function of the threshold $R_l^{max}$ for the transmission heuristic. }
        \label{fig:sensitivity_threshold}
     \end{subfigure}  
     \caption{\small Plot (a): Trade-off between load delivery and wildfire risk, as obtained from the optimal power shut-off with different values of $\alpha$ (blue), the transmission heuristic with different values of $R_{l}^{max}$ (orange) and the area heuristic (green). The stars denote the solutions which we have selected to analyze further in section \ref{op_schemes}. Plot (b): Wildfire risk and load delivery for the optimal power shut-off as the trade-off value $\alpha$ changes. Vertical dashed lines correspond to the $\alpha$ of selected solutions. Plot (c): Wildfire risk and load delivery for the transmission heuristic as the transmission risk threshold $R_{l}^{max}$ is reduced. Vertical dashed lines correspond to the $R_l^{max}$ of selected solutions.}
     \label{fig:sensitivity}
\end{figure*}

The RTS-GMLC system is provided with geographic coordinates and is located in southern California, Nevada, and western Arizona as shown on the map in Fig. \ref{fig:risk_map}.
To associate this test case with realistic values for wildfire risk, we created an artificial wildfire risk map, inspired by a map for Southern California from October 2019 \cite{duginski_2019}. This map is drawn underneath the power lines in Fig. \ref{fig:risk_map}. The wildfire risk levels include low risk (white), medium risk (yellow), high risk (orange), and extreme risk (dark orange). The western region of the map has a very high risk, %
while there is also a region with moderate wildfire risk at the border to Arizona.

We consider the low risk regions to be business-as-usual with wildfire risk $\rho_j=0.0$, and assign wildfire risk values $\rho_j$ of 1.0, 2.0 and 4.0 for the medium, high, and extreme risk zones.
These risk values are then associated with the power system components and their assumed fault probabilities to present the risk of these devices igniting a wildfire.

We assume for simplicity that all generators and buses (i.e., substations) have the same probability $\kappa_{e,j}$ of starting an ignition. Since we are mostly concerned about the relative risk between different components, we assume for simplicity that $\kappa_{e,j}=1$.
For the buses and generators, the wildfire risk coefficient is determined by their locations on the risk map, e.g. bus $i$ in the extreme risk zone $j$ with wildfire risk $\rho_j=4.0$ has a component risk given by $\boldsymbol{R}_i=\kappa_{i,j}\rho_j = 4.0$. 
For the loads, we may consider assigning a higher probability of ignition to a larger load (as a large load is indicative of a larger distribution system). However, other factors than the load size, such as whether a load represents an urban distribution grid with mostly underground cables or a rural community with long high-risk lines, also play a major role in the ignition probability and is not given as part of the available data. For simplicity, we therefore assume $\kappa_{d,j}=1$ (i.e., the ignition probability is assumed to be similar to the generators and buses). As an example, a load in a high risk area with $\rho_j=2.0$ has a component risk of $\boldsymbol{R}_d=\kappa_{d,j}\rho_j=2.0$. Further, we set the weights $w_d=1$ for all loads.

For transmission lines, we assume that the probability of ignition is influenced both by the voltage level and the length of the lines. We assume that the probability of ignition $\kappa_{l,j}$ is uniform along the lines, and that each 10km segment of the line counts as an area $j$ with corresponding wildfire risk $\rho_j$. We choose $\kappa_{l,j}=1.0$ for the $230$ kV lines and $\kappa_{l,j}=2.0$ for the $138$ kV lines, indicating that the lower voltage lines have a higher probability of ignitions relative to the high voltage lines. We then calculate the component risk $\boldsymbol{R}_e$ using \eqref{eq:multiarea}.

\section{Case Study: Numerical Results}
\label{sec:case2}
Using the test case presented above, we now assess the benefit of optimizing the public safety power shut-off under consideration of both wildfire risk and resulting load shedding. To do this, we solve the case study problem using both the optimized power shut-off (OPS) method, the area heuristic (AH) and the transmission heuristic (TH). For each of those models we need to choose some important problem parameters.

For the optimal power shut-off, we need to choose $\alpha$, which determines the trade-off between wildfire risk and load delivery. Choosing  an  optimal  value for $\alpha$ may be possible if we can quantify the economic and societal impact of both wildfires and power blackouts. However, since it is notoriously difficult to obtain accurate estimates of those numbers, we instead run a parameter sweep where we solve the problem for multiple values of $\alpha$ to obtain a Pareto curve. Specifically, we solve the OPS problem with values of alpha ranging from $0$ to $1$ in steps of 0.01.
A grid operator may select any operating point on this curve depending on their wildfire risk tolerance.\\
For the transmission heuristic, we need to select a threshold risk value $R_l^{max}$ beyond which we choose to shut off the line. 
To determine good choices of $R_l^{max}$, we solve the problem with values for $R_l^{max}$ ranging between 70 (higher than the highest risk line) and 0 in steps of 1.  \\
For the area heuristic, the RTS-GMLC case has three predefined regions. We select a risk threshold that disables devices in the high-risk region in the west (corresponding to region 3 in the RTS-GMLC network). 
We also compare our results with just maintaining standard operation, where all components are operational and all load is served.

\subsection{Pareto Front}
To compare the results, we first look at the trade-off between load delivery and wildfire risk obtained with the optimal power shut-off, the transmission heuristic and the area heuristic. Fig. \ref{fig:sensitivity} (a) shows the Pareto front obtained for the optimal power shut-off with varying values of $\alpha$ (blue), the transmission heuristic with varying risk thresholds $R_l^{max}$ (orange), as well as the operating point of the area heuristic, which is only a single point (green). 
From the Pareto front, we observe that a significant reduction in risk can be achieved with no, or very little, load shedding, both with the optimal power shut-off and the transmission heuristic. 
However, the optimal power shut-off method consistently achieves a lower risk level for the same amount of load delivery.
The area heuristic performs much poorer than the other two methods, resulting in both higher load shed and higher wildfire risk. %

To illustrate more explicitly how the solution is impacted by our choice of $\alpha$ and $R_l^{max}$, Fig. \ref{fig:sensitivity} (b) and (c) show the wildfire risk and load delivery as a function of the trade-off parameter $\alpha$ (for the optimal power shut-off)  and the wildfire risk threshold $R_l^{max}$ (for the transmission heuristic). 
In Fig. \ref{fig:sensitivity} (b), the optimal power shut-off shows a sharp drop in risk with a small value of alpha, while the decrease in load delivery is quite small. 
When $\alpha$ reaches $0.25$, the risk is very low (but non-zero) and we are still able to serve approximately 75\% of the system load. As we continue to decrease $\alpha$, the delivered load slowly decreases before it abruptly falls to 0 when $\alpha$ reaches 1.0 (when the problem turns everything off to eliminate all risk, and as a result serve no load).  
For the Transmission Heuristic in Fig. \ref{fig:sensitivity} (c), we see a slightly different behavior. As the transmission line risk threshold decreases from $R_l^{max}=70$ to $20$, the load delivery remains relatively constant (and high), while wildfire risk decreases to less than half of the original value. For $R_l^{max}<20$, both the wildfire risk and the delivered load decrease. With $R_l^{max}=0$, all of the system is de-energized and no load is served.

\begin{figure*}[h!]
    \centering
        \begin{subfigure}[t]{0.3\textwidth}
         \centering
         \textbf{Original Network}
        \makebox[\linewidth][c]{\includegraphics[width=0.95\textwidth]{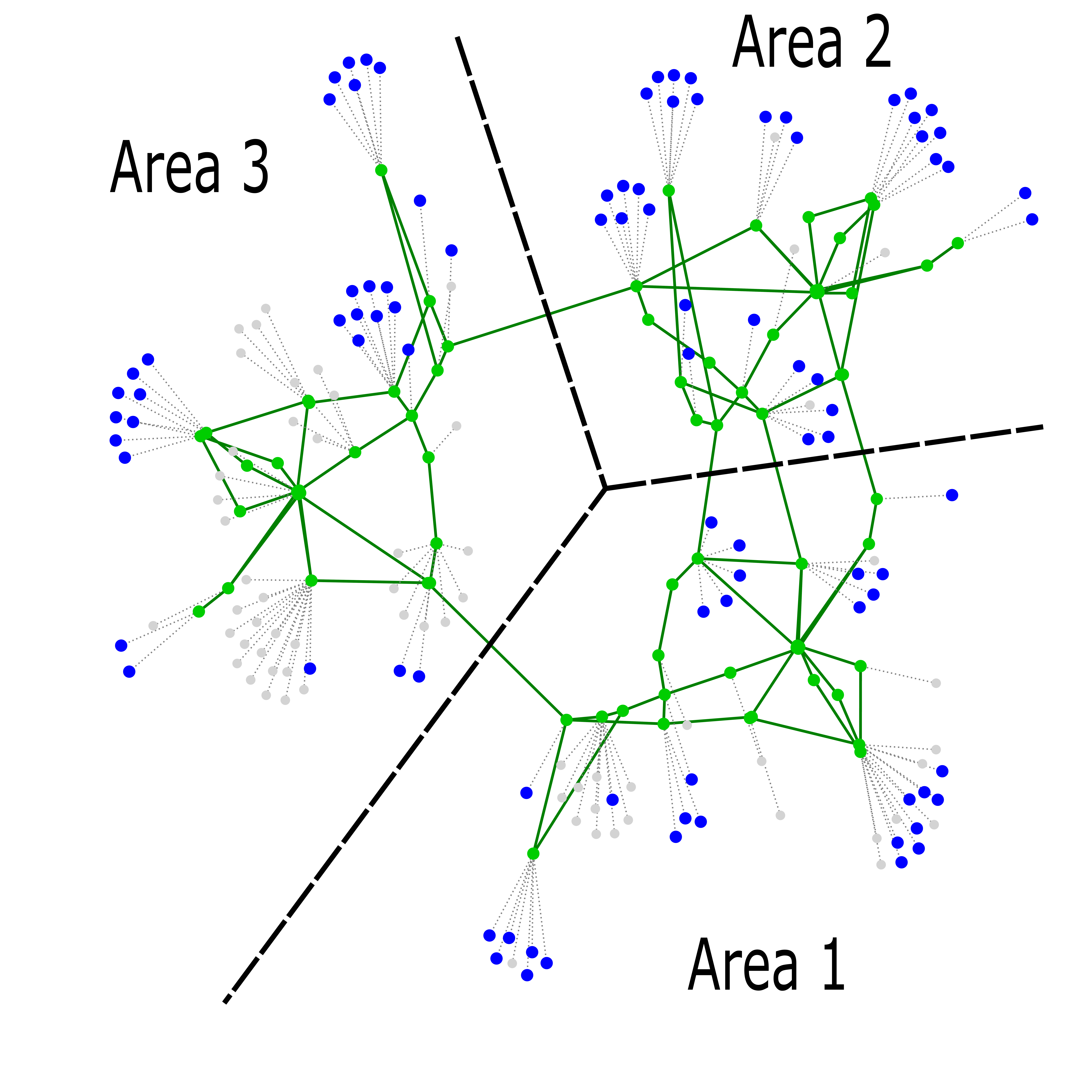}}
         \caption{No shutoff}
         \label{fig:base_network}
     \end{subfigure}
     \begin{subfigure}[t]{0.3\textwidth}
         \centering
            \textbf{ Transmission Heuristic}
            \makebox[\linewidth][c]{\includegraphics[width=0.95\textwidth]{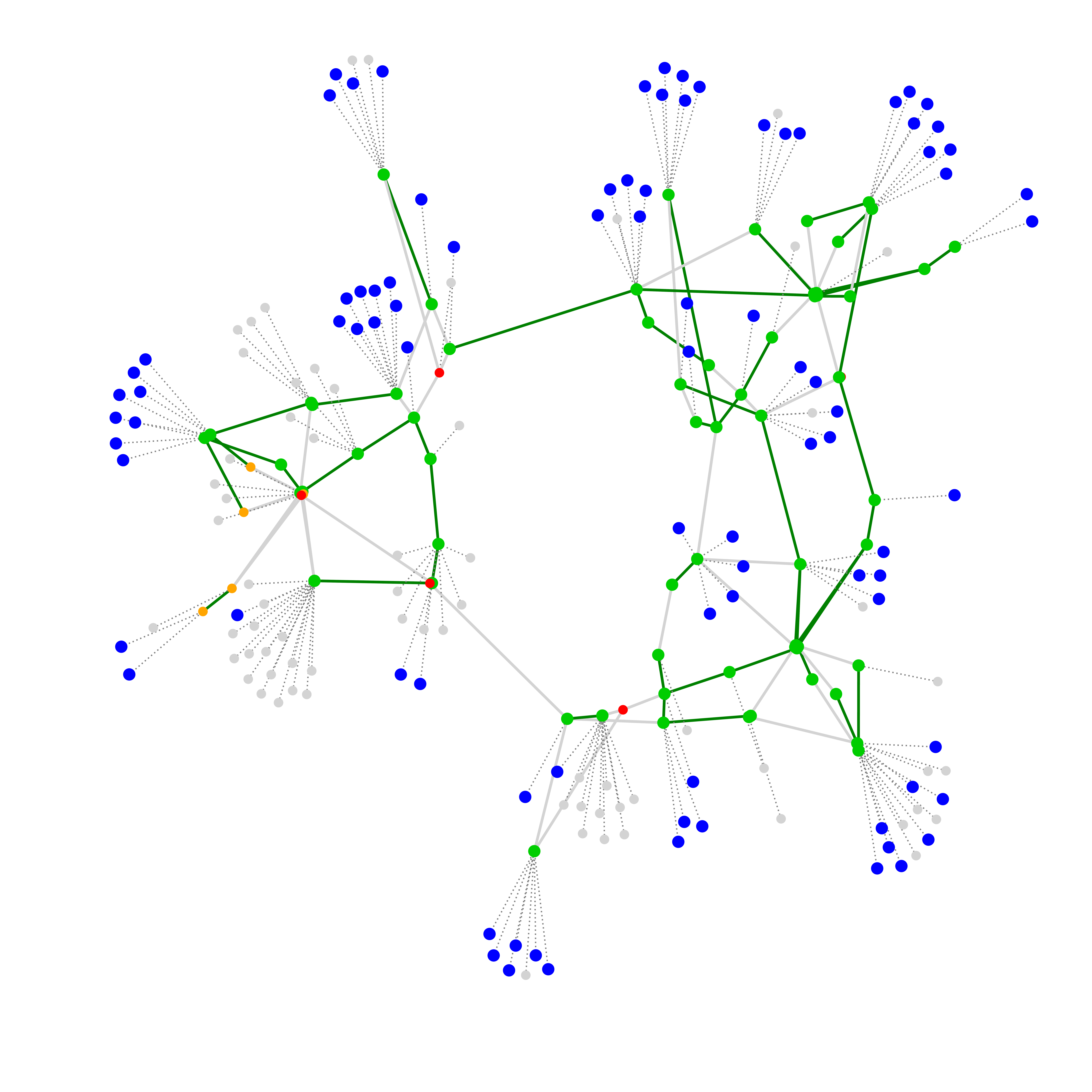}}
            \caption{Medium risk}
         \label{fig:volt_high-risk}
     \end{subfigure}    
     \begin{subfigure}[t]{0.3\textwidth}
         \centering
            \textbf{Optimal Power-Shutoff}
            \makebox[\linewidth][c]{\includegraphics[width=0.95\textwidth]{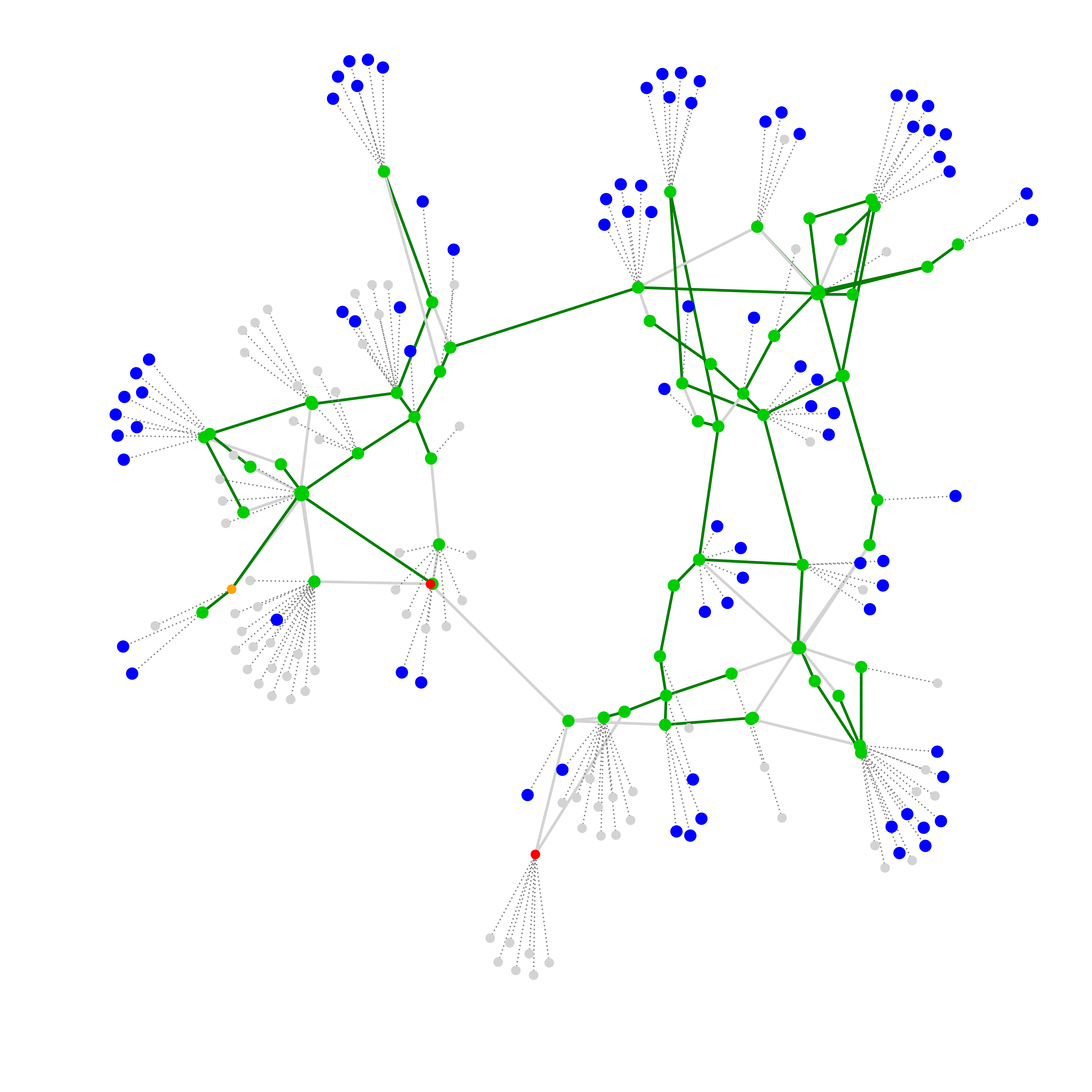}}
            \caption{Medium risk}
         \label{fig:opt_high_risk}
     \end{subfigure}      %
    \par\bigskip
    \begin{subfigure}[b]{0.3\textwidth}
        \centering
            \textbf{Area Shutoff}
            \makebox[\linewidth][c]{\includegraphics[width=0.95\textwidth]{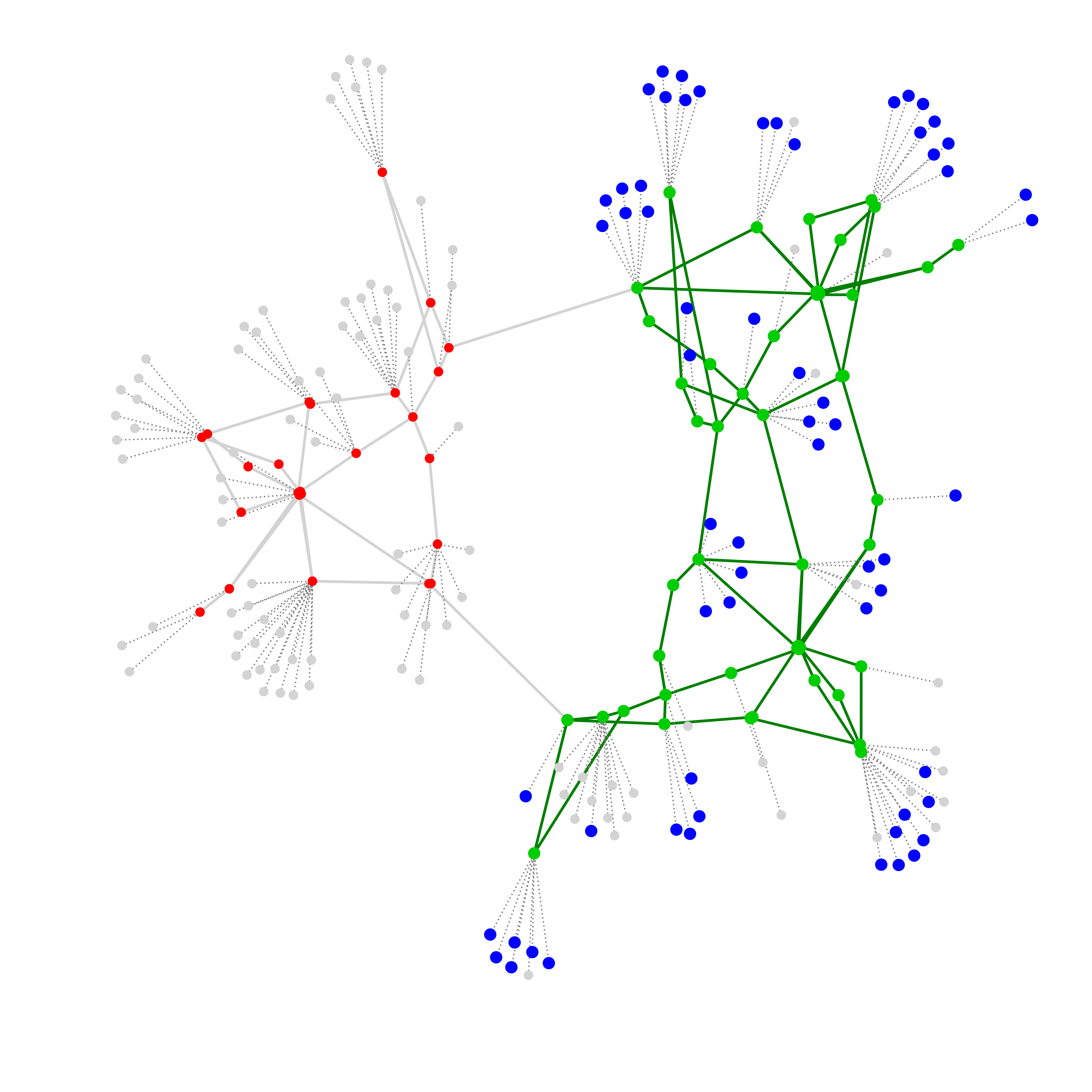}}
            \caption{Regional Shutdown}
        \label{fig:area_risk}
     \end{subfigure}
     \begin{subfigure}[b]{0.3\textwidth}
        \centering
            \textbf{ Transmission Heuristic}
            \makebox[\linewidth][c]{\includegraphics[width=0.95\textwidth]{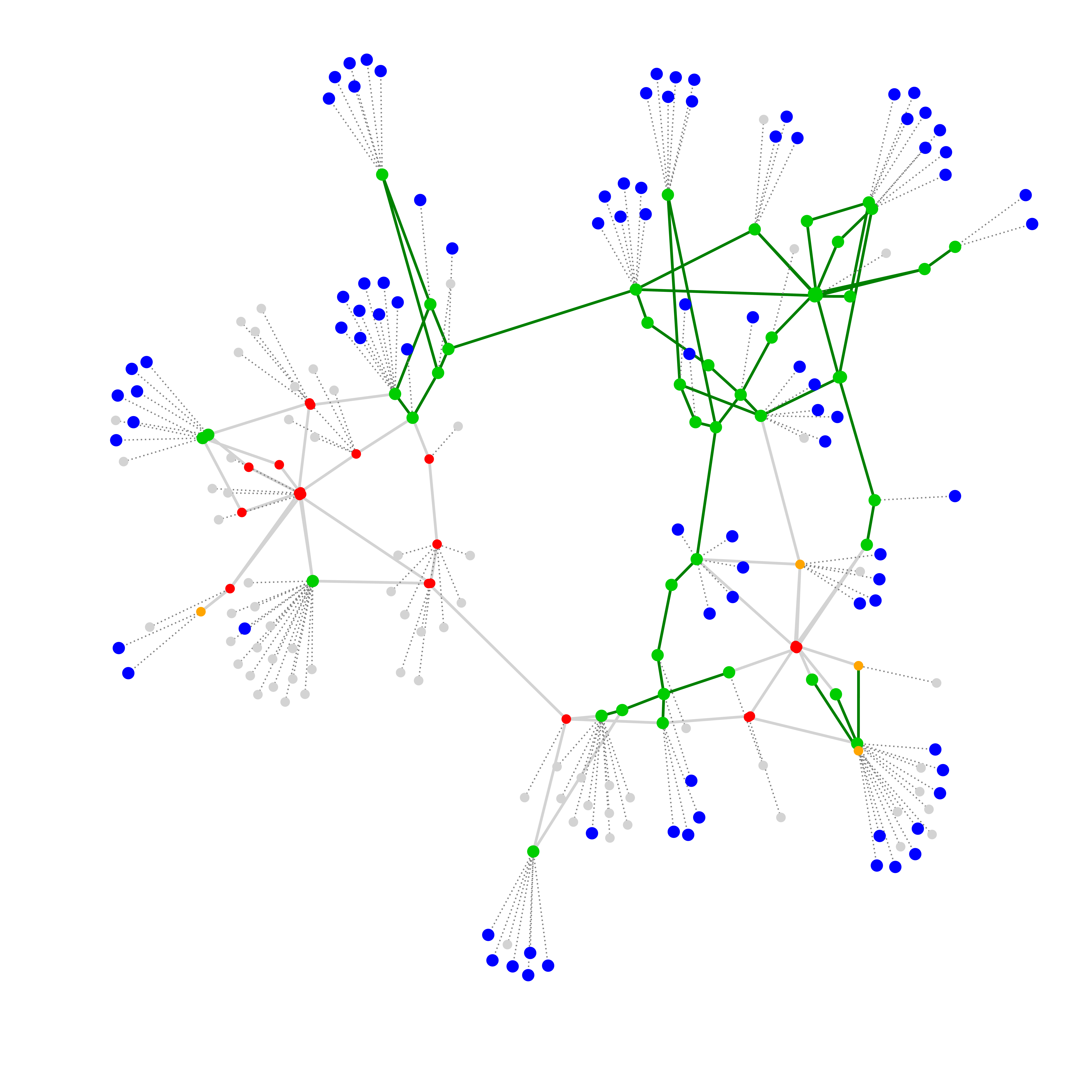}}
            \caption{Low risk}
        \label{fig:volt_low_risk}
     \end{subfigure}    
     \begin{subfigure}[b]{0.3\textwidth}
        \centering
            \textbf{Optimal Power-Shutoff}
            \makebox[\linewidth][c]{\includegraphics[width=0.95\textwidth]{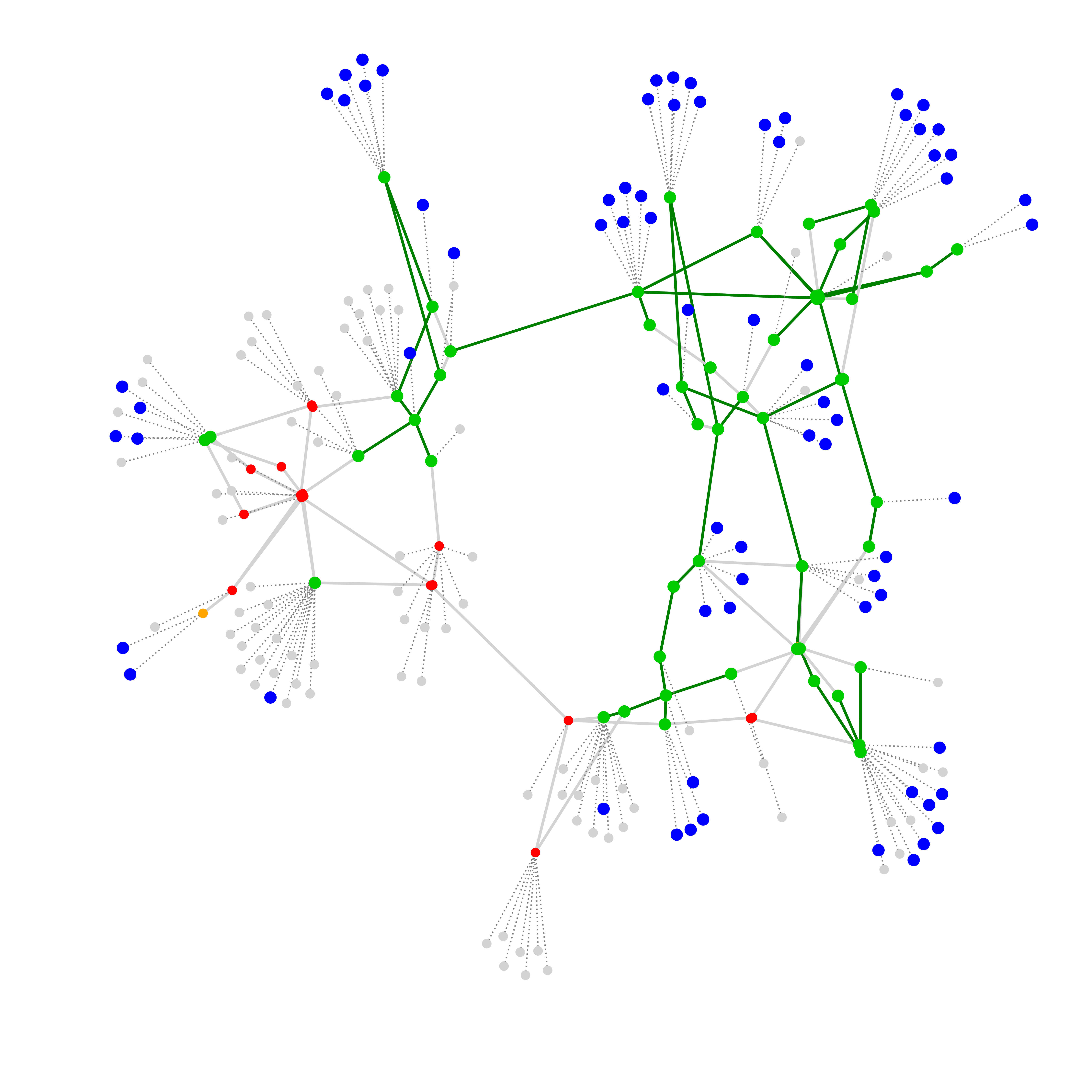}}
            \caption{Low risk}
        \label{fig:opt_low_risk}
     \end{subfigure}
    \caption{\small Plot (a): Original three-area network with no power shutoff applied. Plots (b)-(f): System topology of power shutoff solutions for the medium-risk solutions (top) and low-risk solutions (bottom), with area heuristic (left), transmission heuristic (middle) and optimal power shut-off (right). Inactive components are shown in gray, while active items are shown in green (transmission lines) and blue (generators). Green, orange and red color for the buses indicate delivery of all, some or no load.
    } 
    \label{fig:grids}
\end{figure*} 

\subsection{Operation Schemes} \label{op_schemes}
In Fig. \ref{fig:sensitivity} (a), we select five points for further analysis in the following sections. The selected solutions are indicated by a star, and correspond to a medium and a low risk scenario for the optimal power shut-off and the transmission heuristic, as well as the solution to the area heuristic. For the optimal power shut-off and the transmission heuristic, the same points are indicated by black dashed lines in Fig. \ref{fig:sensitivity} (b) and (c), respectively. 
The medium risk solutions are selected to have a comparable level of wildfire risk, while serving most of the load in the network. These solutions have parameter values of $\alpha = 0.01$ and $R_l^{max} = 18.9$ for the optimal power shut-off and the transmission heuristic, respectively.  
The low risk solutions are selected to have $\approx 10\%$ of the risk of the original network and have parameter values of $\alpha = 0.15$ and $R_l^{max} = 0.0021$.  The Area Heuristic has a regional threshold of  $ R^{max} = 30.0$.
For these scenarios, we plot the resulting network topologies, which are shown in Fig. \ref{fig:grids} (b)-(f). %
Here, inactive components are shown in gray, while active items are shown in green.  Buses with a light-green color fully satisfy load, those in orange have partial load shed, while those in red represent when load is completely shed.  The blue nodes represent active generators, and the attached bus is indicated with a dashed line. Table \ref{tab:operating_risk} lists the total risk, load served, and solve time for those grid scenarios.

\subsubsection{Medium-Risk Operation Schemes}
We first compare the two medium risk scenarios obtained with the optimal power shut-off and the transmission heuristic. As we observe from Table \ref{tab:operating_risk}, these two scenarios have a similar wildfire risk level, but the optimal power shut-off serves more load than the transmission heuristic. Both cases reduce wildfire risk by 50\% relative to standard operation (where all components are energized), at the expense of a reduction in the delivered load from 8550 MW in standard operation to 8540 MW for the optimal power shut-off (10 MW load shed) and 8220 MW for the transmission heuristic (330 MW load shed). We see that the optimal power shut-off serve almost the entire load, while reducing the risk by half.

To better understand the differences between the two solutions, we examine the topologies of the energized grid, which are seen in Fig. \ref{fig:grids} (b) and (c). These figures reveal significant differences in how load is served. The transmission heuristic in Fig. \ref{fig:grids} (b) primarily disables lines in the high-risk western region, and isolates some loads, shown in red. The network in the other regions is almost untouched. In comparison, the optimal power shut-off in Fig. \ref{fig:grids} (c) has  only 10 MW of load shed, and reduces the network to a near-radial, tree-like structure. To serve the load at minimal risk, more components in lower risk zones are disabled, but this is done in a way which allows for almost full load delivery.
It is important to note that the optimal power shut-off removes redundancy in the system in order to reduce risk, and is therefore no longer N-1 secure. While losing this redundancy negatively impacts system reliability, it is preferable to an intentional blackout. 

\begin{table}%
    \caption{\small Total risk and load served for selected operating points}
    \begin{tabular}{lccc}
    \centering
     & \textbf{Total Risk} & \textbf{Load Served} & \textbf{Solve Time} \\[+1pt]
    \hline
    & & & \\[-6pt]
    Standard Operation & 746.2 & 8550 MW & N/A \\[+3pt]
    \textbf{Medium wildfire risk} & & &  \\
    Optimal Power Shutoff & 319.5 & 8540 MW & 0.34 sec  \\
    Transmission Heuristic & 332.8 & 8220 MW & 0.17 sec \\[+3pt]
    \textbf{Low wildfire risk} &  & & \\
    Optimal Power Shutoff & 58.4 & 7210 MW & 0.33 sec  \\
    Transmission Heuristic & 70.8 & 6260 MW & 0.08 sec \\[+3pt]
    Area Heuristic & 186.1 & 5700 MW & 0.04 sec  \\[+1pt]
    \hline
    \end{tabular}
    \label{tab:operating_risk}
\end{table}

\subsubsection{Low-Risk Operation Schemes}
We next examine the two low-risk solutions obtained with the optimal power shut-off solutions and the transmission heuristic, respectively. These low-risk solutions were chosen to have a similar risk level, as can be seen in Table \ref{tab:operating_risk}. We first observe that relative to the medium risk case, the wildfire risk is reduced by an order of magnitude, but at the expense of delivering much less load. The transmission heuristic delivers only 6260 MW load (corresponding to 2290 MW load shed), while the optimal power shut-off achieves 7210 MW load delivery (corresponding to 1340 MW load shed). We conclude that the optimal power shut-off reduces load shed by 25\% relative to the transmission heuristic at comparable risk levels.

When examining the system topologies of the low-risk solutions in Fig. \ref{fig:grids} (e) and (f), %
we observe that both methods shut down almost all components in the high-risk western region. Both approaches also choose to island some regions of the grid, based on the ability of nearby generation to support the load. The main difference between the two solutions is that the optimized strategy is able to prioritize shut-off of components that have a lower impact on load delivery.

We note that the assumption that the grid can operate in multiple islands may or may not be realistic, depending on the generators ability to maintain stable operation, and procedures for resynchronization once the wildfire risk is less acute. However, this result highlights how geographically distributed generation can support local load during large-scale disruptions to the grid.  %

\subsubsection{The value of a targeted approach}
We next take a closer look at the three grid topologies shown in  the bottom part of Fig. \ref{fig:grids},
which from left to right represent an increasingly targeted approach to public safety power shutdown. Our least targeted approach is the area heuristic in Fig. \ref{fig:grids} (d), which shuts down an entire region based on a simple threshold. The transmission heuristic in Fig. \ref{fig:grids} (e) is more targeted because it considers the risk values of individual lines. Finally, the optimal power shut-off in Fig. \ref{fig:grids} (f) is the most comprehensive approach, as it considers both the risk of individual components as well as their impact on the system's ability to serve the load. Comparing the grid topologies and the values for load delivery and wildfire risk in Table \ref{tab:operating_risk}, we observe that including mode detail consistently enables a solution with lower risk \emph{and} increased load delivery.

\begin{table}
    \centering
    \caption{\small Total wildfire risk for different types of devices in the solutions obtained with the optimal power shut-off problem.}
    \begin{tabular}{ lccccc}
         \textbf{Optimal Power Shut-Off}& \textbf{Total} & \textbf{Bus} & \textbf{Line} & \textbf{Gen.} & \textbf{Load} \\[+1pt] 
         \hline
         &  &  & &  & \\[-6pt]
         Medium Wildfire Risk & 319.5 & 52.0 & 181.7 & 26.0 & 59.9 \\
         Low Wildfire Risk & 58.4 & 15.0 & 12.6 & 13.0 & 17.8 \\[+1pt]
    \hline
    \end{tabular}
    \label{tab:opt_risk}
\end{table}

\begin{figure*}[t]
    \centering
        \begin{subfigure}[t]{0.45\textwidth}
            \centering
            \includegraphics[width=0.90\textwidth]{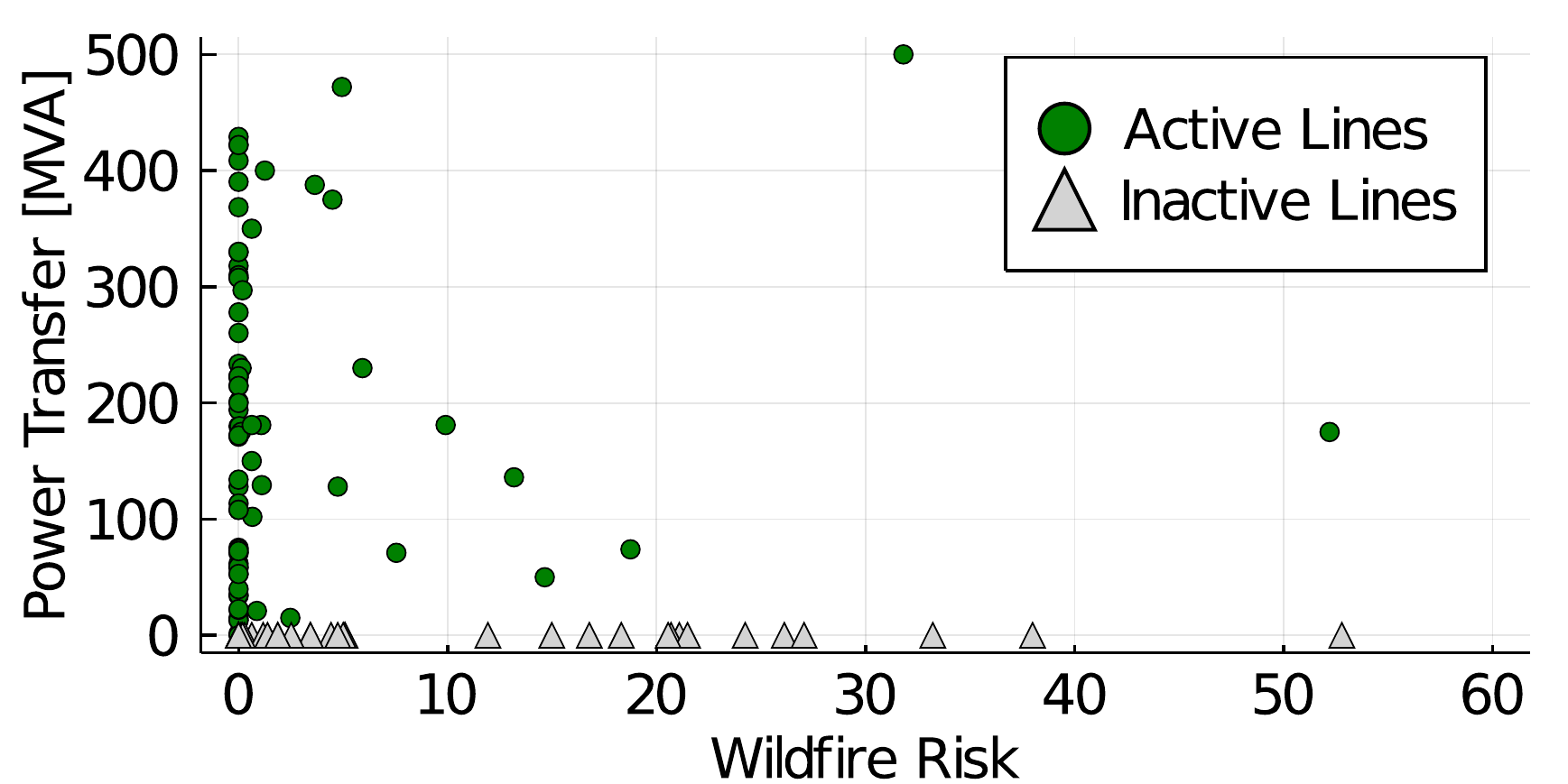}
            \caption{\small \textbf{Medium risk operation.} }
            \label{fig:scatter_opt_high_risk}
     \end{subfigure}
     \begin{subfigure}[t]{0.45\textwidth}
        \centering
        \includegraphics[width=0.90\textwidth]{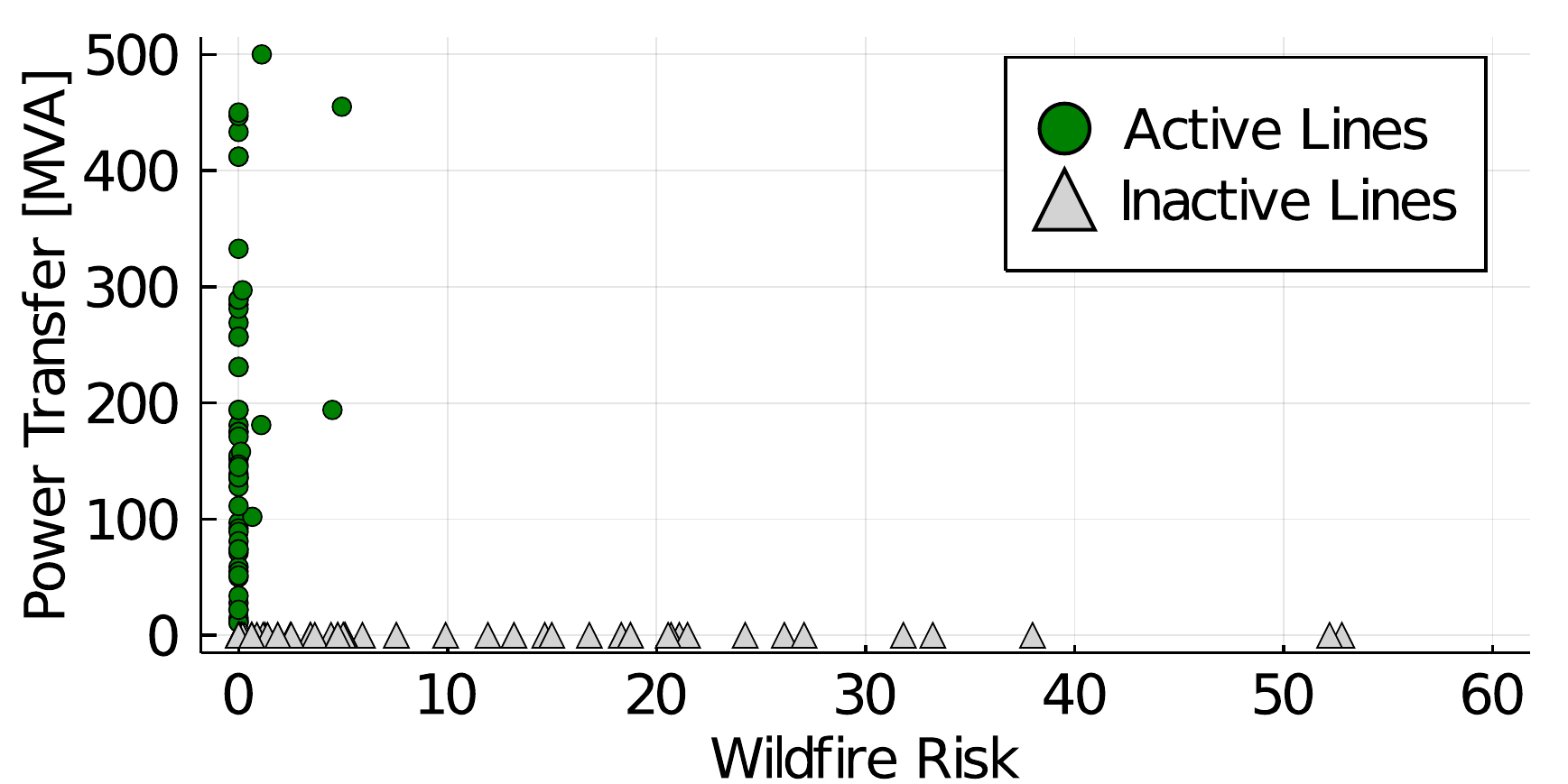}
        \caption{\small \textbf{Low risk operation.} }
        \label{fig:scatter_opt_low_risk}
     \end{subfigure}  
     \caption{\small Scatter plot showing the wildfire risk (horizontal axis) and the power transferred (vertical axis) for each transmission line.}%
     \label{fig:active_lines}
\end{figure*}

\subsection{Optimal Power Shut-Off Solutions}
To better understand the optimal power shut-off solutions, we investigate how the total risk is distributed among the different component types. Table \ref{tab:opt_risk} shows the risk is associated with each component type in the medium and low risk solutions. The two cases differ mainly in how much wildfire risk is associated with transmission lines. Transmission lines contribute 57\% of the total risk in the medium-risk case, compared with only 20\% in the low-risk case. 

To understand transmission line risk in more detail, we plot the risk coefficient $\boldsymbol{R}_l$ for each line against the amount of power carried by this line, $P_{l,i,j}^L$. The result is shown in Fig. \ref{fig:active_lines}, with the  medium risk solution on the left and the low risk solution on the right. The energized transmission lines are represented by blue dots and de-energized lines are plotted in orange (note that $P_{l,i,j}^L=0$ for de-energized lines). %
From these plots, we first observe that the risk values of the lines vary between 0 and 53, with most lines concentrated at lower values. We notice that the low-risk solution turns off essentially all high-risk power lines, while the medium risk solution allows even some of the highest risk lines continue to operate. Further, we see that the threshold for when the lines are turned off is not given by the absolute risk value of the line (as it is in the threshold heuristic), but is rather a function of both the risk value of the line and the amount of power transferred by the line. These results demonstrate how the optimal power shut-off problem allows high risk lines to operate if they transfer a significant amount of power, and thus contribute significantly to increase the amount of delivered load. %

\section{Conclusions and Future Work}
\label{sec:conclusions}
This work proposes a new optimization model to minimize wildfire risk due to electric power system components, while maintaining electricity supply to as many customers as possible.
The first contribution is to devise a principled strategy to assign risk values to electric components. Specifically, we propose to utilize extensions of existing models for wildfire risk to assign risk values to individual electrical components, but extend those risk values to account for information about elevated or reduced ignition probability based on component characteristics. This allows us to assess the reduction in wildfire risk that is achieved by de-energizing the component. Our second contribution is to formulate the optimal power shut-off problem that utilizes these risk values. This problem is similar to an optimal power flow with additional decision variables to represent whether or not a component is energized. The objective is to minimize wildfire risk, while maximizing load delivery. 
This proposed approach is demonstrated on the RTS-GMLC test case, which is combined with data from a wildfire risk map. The optimal power shut-off problem is compared against two heuristic decision models which disable specific components based on their wildfire risk value alone (i.e., without consideration of the impact on load shed). %
We make the three significant observations:
\begin{itemize}
    \item The optimal power shut-off problem, which considers both the wildfire risk value and the impact on the delivered load, is able to serve more load at lower risk values compared with the threshold-based heuristic.
    \item This is achieved by allowing highly loaded lines (which are important to maintain load delivery) to continue to operate at higher wildfire risk values, while less loaded lines are shut-off at lower risk values.
    \item The minimal risk network topology has a tree-like structure, and the system is sometimes split into islands.
\end{itemize}

Given those observations, we consider our model as a first, important step to  better understand how power system operations can be modified to minimize wildfire risk, at minimal disruption to customers.

However, there are several important avenues for future work. 
We believe that it is important to develop a better and more detailed model of the wildfire risk arising from electric power infrastructure, which reflects current weather conditions and allows us to assess the risk reduction achieved through different wildfire mitigation measures. This would allow for better cost-benefit analysis. 
We also believe that it is necessary to consider 
multi-period optimal power flow problem which uses data for electric load and renewable energy sources to identify how to minimize wildfire risk and maximize load delivery throughout several days. Choosing a single fixed topology, rather than constantly re-configuring the network based on changing risks, is necessary since de-energized lines require inspection before they can be safely re-energized. This would also naturally fit with the modelling of energy storage and local generation, which could allow certain regions to operate as locally run microgrids for short periods of time.

Another important consideration is the impact of additional electric faults that occur after the power shut-off is already in place. Our optimal power shut-off leads to a network topology that is no longer N-1 secure, and as a result, additional faults have significant impact on reliability and expected load delivery. In future work, we would like to consider the impact of N-1 contingencies in addition to the power shut-offs.
Extensions to account for more detailed distribution grid modelling are also important, as distribution grid lines cause more ignitions. 

Finally, our model, which currently optimizes real-time operations, could be extended to consider longer-term planning decisions such as 
the construction of new distributed generation to mitigate the impact of public safety power shut-offs or identification of the most critical lines for wildfire mitigation (e.g., to prioritize installation of new protection or monitoring devices, or schedule vegetation management). 

Overall, we believe that this paper is an important step towards a better understanding of the relationship of wildfire risk and electric grid operation. The cost associated with mitigating this risk will be essential to improve power grid operation in wildfire-prone areas, where the risk of wildfires and the impact of power outages creates extremely challenging trade-offs.

\bibliographystyle{IEEEtran}
\bibliography{references}

\end{document}